\documentclass[twocolumn,amsmath,showpacs,superscriptaddress,prb,aps]{revtex4-2}
\usepackage{graphicx}
\usepackage{bm}
\usepackage{xcolor}   
\usepackage{amssymb} 
\usepackage{amsmath}
\usepackage[utf8]{inputenc}
\usepackage{hyperref}
\usepackage{siunitx}
\hypersetup{
	colorlinks=true,
	linkcolor=blue,
	filecolor=magenta,
	urlcolor=cyan,
}

\usepackage[normalem]{ulem}

\begin{document}
	
\title{Phonon-Assisted Tunneling through Quantum Dot Systems Connected to Majorana Bound States}
	
\author{Levente M\'{a}th\'{e}}
\affiliation{Center of Advanced Research and Technologies for Alternative Energies, National Institute for R \& D of Isotopic and Molecular Technologies, 67-103 Donat, 400293 Cluj-Napoca, Romania}
\affiliation{Faculty of Physics, Babeș-Bolyai University, 1 Kogălniceanu, 400084 Cluj-Napoca, Romania}

\author{Zoltán Kovács-Krausz}
\affiliation{Department of Physics, Institute of Physics, Budapest University of Technology and Economics, Műegyetem rkp. 3, H-1111 Budapest, Hungary}

\author{Ioan Botiz}
\affiliation{Faculty of Physics, Babeș-Bolyai University, 1 Kogălniceanu, 400084 Cluj-Napoca, Romania}
\affiliation{Interdisciplinary Research Institute in Bio-Nano-Sciences, Babeș-Bolyai University, 42 Treboniu Laurian, 400271 Cluj-Napoca, Romania}

\author{Ioan Grosu}
\affiliation{Faculty of Physics, Babeș-Bolyai University, 1 Kogălniceanu, 400084 Cluj-Napoca, Romania}

\author{Khadija El Anouz}
\affiliation{Laboratory of R \& D in Engineering Sciences, Faculty of Sciences and Techniques Al‑Hoceima, Abdelmalek Essaadi University, Tetouan 93000, Morocco}

\author{Abderrahim El Allati}
\affiliation{Laboratory of R \& D in Engineering Sciences, Faculty of Sciences and Techniques Al‑Hoceima, Abdelmalek Essaadi University, Tetouan 93000, Morocco}

\author{Liviu P. Z\^{a}rbo}
\email[Corresponding author: ]{liviu.zarbo@itim-cj.ro}
\affiliation{Center of Advanced Research and Technologies for Alternative Energies, National Institute for R \& D of Isotopic and Molecular Technologies, 67-103 Donat, 400293 Cluj-Napoca, Romania}

\date{\today}
	
\begin{abstract}
We theoretically analyze phonon-assisted tunneling transport in a quantum dot side connected to a Majorana bound state in a topological superconducting nanowire. We investigate the behavior of the current through the dot, for a range of experimentally relevant parameters, in the presence of one long-wave optical phonon mode. We consider the current-gate voltage, the current-bias voltage and the current-dot--Majorana coupling characteristics under the influence of the electron--phonon coupling.
In the absence of electron--phonon interaction, the Majorana bound states suppress the current when the gate voltage matches the Fermi level, but the increase in the bias voltage counteracts this effect.
In the presence of electron--phonon coupling, the current behaves similarly as a function of the renormalized gate voltage. As an added feature at large bias voltages, it presents a dip or a plateau, depending on the size of the dot--Majorana coupling. 
Lastly, we show that the currents are most sensitive to, and depend non-trivially on the parameters of the Majorana circuit element, in the regime of low temperatures combined with low voltages. Our results provide insights into the complex physics of quantum dot devices used to probe Majorana bound states.
\end{abstract}

\keywords{quantum dot; Majorana bound states; electron--phonon interaction; optical phonon; quantum transport}
\pacs{}
\maketitle
	
\section{INTRODUCTION}
\label{sec:I}

Majorana bound states (MBSs) are zero-energy excitations in topological materials known to form a potential platform for solid state quantum computation due to their non-Abelian statistics~\cite{Kitaev2001,Kitaev2006,Nayak2008,Alicea2012,Aguado2017}.
Previous theoretical works~\cite{Lutchyn2010,Oreg2010} considered devices based on semiconducting nanowires realized from InAs or InSb with strong spin--orbit coupling located in the proximity of $s$-wave superconductors (SCs). The~latter were threaded by an external magnetic field in order to drive the nanowire into its topological superconducting phase by creating MBSs. Such theoretical proposals were further confirmed via  experiments~\cite{Mourik2012}. Other theoretical works proposed experimental setups realized from topological insulators~\cite{Fu2008}, magnetic nanoparticles on SCs~\cite{Choy2011}, nanomagnets~\cite{Kjaergaard2012} and $p$-wave SCs~\cite{Read2000,Stone2004} to create~MBSs.

A minimal setup to probe MBSs in topological superconducting nanowires (TSNWs) requires the coupling of the nanowire to a quantum dot (QD), which introduces regular fermionic degrees of freedom~\cite{Flensberg2011,Liu2011}. The~presence of MBSs requires the conductance to take the value of $e^2/2h$, which is measured through the QD via normal leads~\cite{Liu2011}.
Several theoretical designs based on either single QDs--TSNWs or double QD interferometer--TSNW setups~\cite{Zou2023} have been considered in order to probe the MBSs via transport properties such as (thermal)~\cite{Leijnse2014,Lopez2014,Khim2015,Chi2020a,Grosu2023,Zou2023} conductance~\cite{Cao2012,Lee2013,Vernek2014,Dessotti2014,Stefanski2015,Zeng2016b,Calle2020,Ricco2020,Majek2021,Smirnov2021,Gong2022a,Gong2022b,Jiang2022,Zou2023}, current noise~\cite{Cao2012,Lu2012,Chen2014a,Lu2014,Smirnov2022} and Josephson current~\cite{Lee2016,Feng2023}.
Details on the experimental detection of MBSs via transport characteristics measurements have been reported elsewhere~\cite{Mourik2012,Finck2013,Lee2014,Chen2017,Zhang2017,Lutchyn2018,Deng2016,Sherman2017,Deng2018}.
The photon-assisted transport properties of QD--MBS setups have been studied in the literature both theoretically~\cite{Tang2015,Chi2020,He2021,Ricco2022a,Ricco2022b} and experimentally~\cite{Zanten2020}.

Over the last few years, the~effect of optical phonons on the transport properties of QD--MBS systems has attracted great attention~\cite{Dai2019,Wang2020,Wang2021,Wang2021b,Wang2021c,Mathe2022}. 
The phonon-assisted transport properties of QDs coupled to MBSs have been studied in Refs.~\cite{Wang2021b,Wang2021c} in order to establish the connection between the electrical current and heat generation in such systems. 
In a recent study, we analyzed the phonon-assisted transport properties in a QD connected to a Majorana ring structure~\cite{Mathe2022}. We found the periodicity of zero-temperature linear conductance, as~a function of threading magnetic flux phase, to~be independent of the electron--phonon interaction (EPI), as~well as of changes in QD energy and finite values of the QD--MBS couplings when the Majorana wave functions do not~overlap.
\begin{figure}[ht]
	\includegraphics[width =1\linewidth]{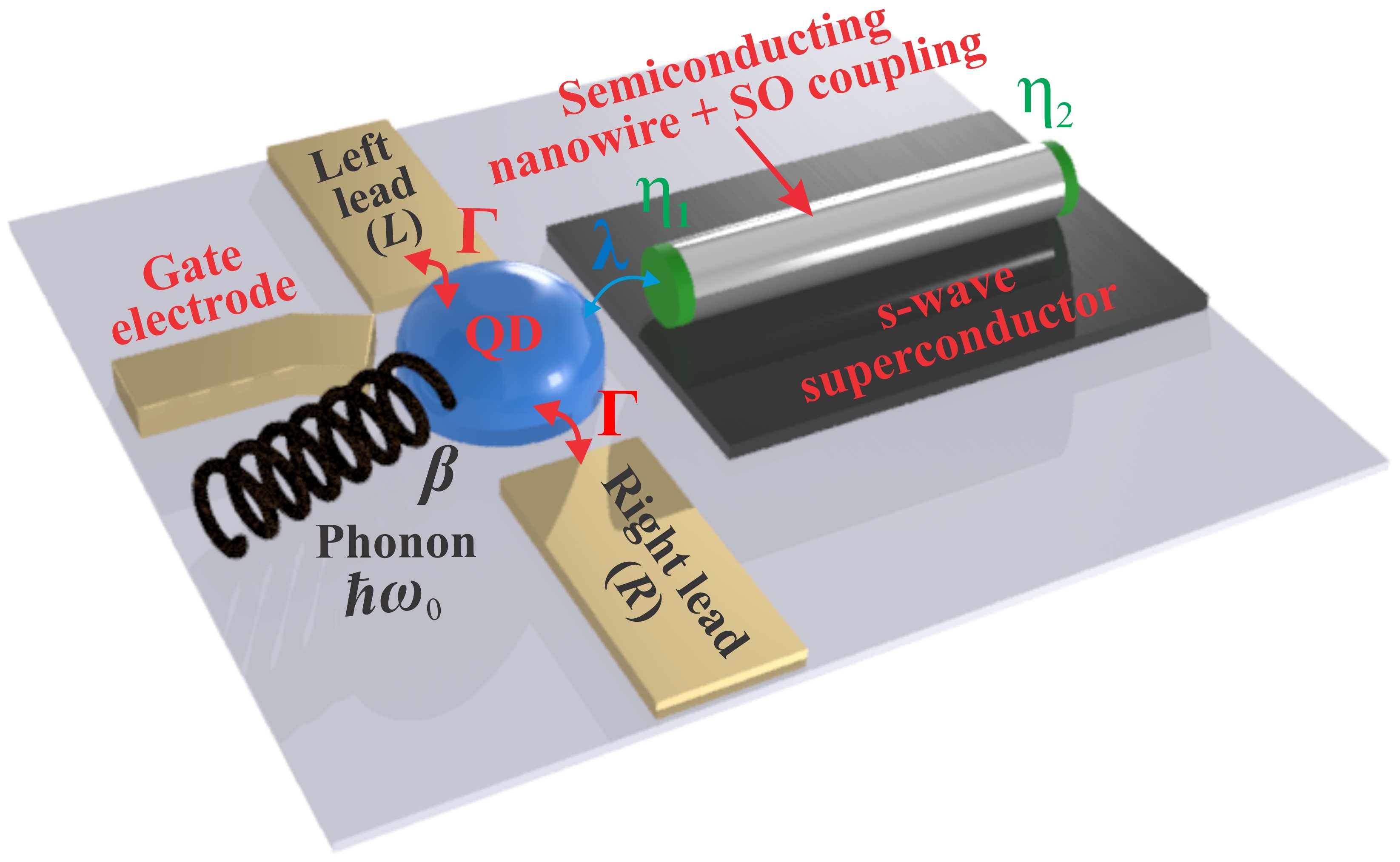}
	\centering
	\caption{Schematic representation of a QD connected to a MBS located at one of the ends of a TSNW. Here, $\eta_{1}$ and $\eta_2$ denote the Majorana operators corresponding to the two MBSs in the TSNW. The~dot is coupled to two normal leads with coupling strength $\Gamma$. The~electron in QD interacts with a single long-wave optical phonon mode of frequency $\omega_0$. The~notations $\lambda$ and $\beta$ represent the QD--MBS and electron--phonon coupling strengths, respectively.}
	\label{fig1}
\end{figure}

In this work, we study the phonon-assisted transport properties of a QD coupled to one MBS located at one of the ends of a TSNW. We measure the current through the QD via metallic leads. The~QD interacts with a single long-wave optical phonon mode. We treat the EPI within a canonical transformation which leads to the renormalization of QD energy, dot-leads and dot--MBS couplings. Here, we consider the effect of EPI strength and temperature on the QD--leads and QD--MBS couplings, which is usually neglected in the literature~\cite{Wang2021b,Wang2021c}.
Under such conditions, we discuss in detail the transport characteristics of the considered system for unhybridized and hybridized~MBSs.

The article is organized as follows. In~Section~\ref{sec:II}, we present the theoretical model used by us, and perform a canonical transformation on the system Hamiltonian to eliminate the EPI. We then calculate the tunneling current using the nonequilibrium Green's function method. We show and discuss the results in Section~\ref{sec:III}. Finally, we present our conclusions and discuss the significance of our main results in Section~\ref{sec:IV}.

\section{Theory}
\label{sec:II}
We consider a QD connected to two normal leads and to one of the ends of a TSNW via a MBS, as~it is shown in Figure~\ref{fig1}. The~normal leads allow the measuring of a current through the QD. The~dot energy is tuned by the gate voltage $V_g$ applied to the gate electrode. The~interaction between the QD electron and the single long-wave optical phonon mode leads to phonon-assisted transport. The~Zeeman energy, $V_Z$, the~largest energy scale in the system, is created by the applied magnetic field which drives the nanowire into the topological superconducting phase when the relation $V_Z > \sqrt{\Delta^2 + \mu^2}$ is fulfilled. Here, $\Delta$ and $\mu$ represent the SC energy gap and nanowire chemical potential, respectively.
The considered system is described by the Hamiltonian~\cite{Liu2011,Zhu2003,Chen2005,Ramos2018,Mathe2022}:
\begin{equation}
	H=H_{\text{leads}}+H_{\text{MBS}}+H_{\text{ph}}+H_{\text{QD}}+H_{\text{tun}}.
	\label{eq1}
\end{equation}
The~Hamiltonian $H_{\text{leads}}$ in Eq.~\eqref{eq1} models the noninteracting electrons in both leads,
\begin{equation}
	H_{\text{leads}}=\sum_{\gamma,k}\varepsilon_{\gamma k}\,c_{\gamma k}^{\dagger}c_{\gamma k},
	\label{eq2}
\end{equation}
where $c_{\gamma k}^{\dagger}$ ($c_{\gamma k}$) is the creation (annihilation) operator for an electron with momentum $k$ in the left ($\gamma = L$) and right ($\gamma = R$) leads.
Therefore, $\varepsilon_{\gamma k}=\varepsilon_{k}-\mu_{\gamma}$ represent the single-particle energies and the chemical potential $\mu_\gamma$. In the following, the~leads are at the same temperature ($T_\gamma=T$).
The next term in Eq.~\eqref{eq1}, $H_{\text{MBS}}$, describes the interaction between MBSs,
\begin{equation}
	H_{\text{MBS}}=i \varepsilon_M \eta_1 \eta_2,
	\label{eq3}
\end{equation}
where $\varepsilon_M \propto e^{-L/\xi}$ is the Majorana overlap energy with the TSNW length ($L$) and superconducting coherence length ($\xi$).
Here, $\eta_1$ and $\eta_2$ are the Majorana operators related to the two MBSs located at the opposite ends of the~TSNW.
The third term in Eq.~\eqref{eq1}, $H_{\text{ph}}$, models the longitudinal optical phonon mode,
\begin{equation}
	H_{\text{ph}}=\hbar \omega_0 a^\dagger a,
	\label{eq4}
\end{equation}
with the energy $\hbar \omega_0$. Here, $a^{\dagger}$ and $a$ are the phonon creation and annihilation operators. 
The QD Hamiltonian, $H_\text{QD}$, is given by
\begin{equation}
	H_\text{QD}=\varepsilon_{d}d^{\dagger}d +\beta (a+a^\dagger)d^\dagger d,
	\label{eq5}
\end{equation}
where $\varepsilon_{d}$ is the QD energy and $d^{\dagger}(d)$ is the creation (annihilation) operator for an electron in the QD. The~EPI is modeled by the second part of Eq.~\eqref{eq5} where $\beta$ is the electron--phonon coupling strength.
The last term in Eq.~\eqref{eq1}, $H_\text{tun}$, represents the tunneling~Hamiltonian,
\begin{equation}
	H_\text{tun}=(\lambda d - \lambda^* d^\dagger)\eta_1 + \sum_{\gamma,k}\big (V_{\gamma k}c_{\gamma k}^{\dagger}d +V^*_{\gamma k}d^{\dagger}c_{\gamma k}\big ),
	\label{eq6}
\end{equation}
where the first term in Eq.~\eqref{eq6} describes the coupling of strength $\lambda$ between the QD and the MBS $\eta_1$ located at one of the end of TSNW. 
The second component of \mbox{Eq.\eqref{eq6}} refers to the coupling between the QD and the lead $\gamma$ and~is characterized by the $V_{\gamma k}$ tunneling amplitude.
For further calculations, the~Majorana operators $\eta_1$ and $\eta_2$ in \mbox{Eqs.~\eqref{eq3} and~\eqref{eq6}} will be replaced with regular fermionic operators $\eta_1=(f^\dagger +f)/ \sqrt{2}$ and $\eta_2=i(f^\dagger -f)/ \sqrt{2}$.
In order to explore the transport properties of the system, we bias the QD as $\mu_L=-\mu_R=eV/2$, while the SC is grounded, i.e.,~$\mu_S = 0$. We also restrict our calculations to the wide-band limit~\cite{Jauho1994} for a symmetrically coupled QD--lead system displaying an electron-hole symmetry, i.e.,~$\Gamma_{\gamma}^{e}=\Gamma_{\gamma}^{h}=\Gamma_{\gamma}=\Gamma$, where $\Gamma_{\gamma}^{e(h)} = 2\pi \sum_{k} |V_{\gamma k}|^2 \delta (\varepsilon \mp \varepsilon _{\gamma k})$ is the coupling between the dot and the lead $\gamma$ for electrons (holes).
Furthermore, we consider the relatively weak electron--phonon coupling limit~\cite{Zhu2003}, by~employing the nonequilibrium Green's function technique~\cite{Jauho1994,Sun2000,Michalek2013} in the subgap regime $|eV|<\Delta$. In~this case, at~finite temperature the current takes the form~\cite{Mathe2022}:
\begin{equation}
	I = \frac{ie}{2h}\Gamma \int d\varepsilon \big[f_{L}^{e}(\varepsilon) - f_{R}^{e}(\varepsilon)\big] [\mathbf{G}_d^>(\varepsilon) - \mathbf{G}_d^<(\varepsilon)]_{11}.
	\label{eq7}
\end{equation}
Here, $f_{\gamma}^{e}(\varepsilon)$ represents the Fermi--Dirac distribution function for electrons in lead $\gamma$ and $\mathbf{G}_{d}^{<(>)}(\varepsilon)$ is the lesser (greater) Green's function matrix of the QD in Nambu space~\cite{Hwang2015,Bocian2015}. Next, we set $k_B = \hbar = 1$.
The corresponding lesser and greater Green's functions, appearing in Eq.~\eqref{eq7}, are determined by using a canonical transformation $\bar H = e^S H e^{-S}$ with $S=(\beta/\omega_0)d^{\dagger} d (a^{\dagger} - a)$ which aims to eliminate the electron--phonon coupling term in the Hamiltonian given by Eq.~\eqref{eq1}~\cite{Zhu2003,Chen2005,Swirkowicz2008,Mathe2018,Mathe2022}.
Thus, the~transformed Hamiltonian becomes $\bar H = \bar H_{\text{El}} + H_{\text{ph}}$ where the electron term is
\begin{equation}
	\bar H_{\text{El}} = H_{\text{leads}} + H_{\text{MBS}} +\tilde H_{\text{QD}} + \bar H_{\text{tun}},
	\label{eq8}
\end{equation}
with
\begin{equation}
	\bar H_\text{tun}= \frac{1}{\sqrt{2}}(\tilde{\lambda} d - \tilde{\lambda}^* d^{\dagger})(f + f^{\dagger})
	+ \sum_{\gamma,k}\big (\tilde{V}_{\gamma k}c_{\gamma k}^{\dagger}d +\tilde{V}^*_{\gamma k}d^{\dagger}c_{\gamma k}\big ).
	\label{eq9}
\end{equation}
Here, $\tilde H_\text{QD}=\tilde {\varepsilon}_{d}d^{\dagger}d$ and the renormalized dot energy $\tilde {\varepsilon}_d = \varepsilon_d - g \omega_0$, with~$g$ being equal to $g=(\beta/\omega_0)^2$.
Furthermore, the~QD--leads and QD--MBS couplings become renormalized as $\tilde{V}_{\gamma k} = V_{\gamma k} X$ and $\tilde{\lambda} = \lambda X$ with $X = \exp(-(\beta/\omega_0)(a^{\dagger}-a))$. 
Here, we apply the approximation $X \approx \langle X \rangle = \exp(-g(N_{\text{ph}}+\frac{1}{2}))$~\cite{Chen2005}, where $N_{\text{ph}}=1/(e^{\omega_0/T}-1)$ is the Bose--Einstein distribution function. This approximation holds if $V_{\gamma k}, \, \lambda \ll \text{min}(\beta,\Delta)$ or $\beta \ll \text{min}(V_{\gamma k},\lambda,\Delta)$~\cite{Chen2005,Zhang2012,Dai2019}. 
The lesser and greater Green's functions are given by
\begin{equation}
	\begin{aligned}
		\mathbf{G}_d^<(\varepsilon)&=\sum_{l=-\infty}^{\infty}\mathcal{L}_l \mathbf{ \tilde G}_d^<(\varepsilon+l\omega_0),\\
		\mathbf{G}_d^>(\varepsilon)&=\sum_{l=-\infty}^{\infty}\mathcal{L}_l \mathbf{ \tilde G}_d^>(\varepsilon-l\omega_0),\\
	\end{aligned}\label{eq10}
\end{equation}
where $\mathcal{L}_l=e^{-g(2N_{\text{ph}}+1)}e^{l\omega_0/(2T)}I_l(2g\sqrt{N_\text{ph}(N_{\text{ph}}+1)})$ is the Franck--Condon factor at finite temperature~\cite{Chen2005}.
Here, $I_l(z)$ is the \textit{l}th order modified Bessel function of the first kind. Note that $\mathcal{L}_l$ becomes $\mathcal{L}_l = e^{-g} g^l/l!$ for $l \ge 0$, while $\mathcal{L}_l = 0$ for $l<0$ at $T=0$.
The dressed lesser (greater) Green's function $\mathbf{\tilde G}_d^{<(>)}$ is calculated by employing the Keldysh equation $\mathbf{\tilde G}_d^{<(>)}=\mathbf{ \tilde G}_d^r \mathbf{\tilde\Sigma}^{<(>)}\mathbf{\tilde G}_d^a$ with the use of corresponding lesser (greater) self-energy $\mathbf{\tilde\Sigma}^{<(>)}$. Thus, the~current given in relation~\eqref{eq7}, at~finite temperature, reads~\cite{Mathe2022}\vspace{-12pt}
\begin{widetext}
	\begin{equation}
		\begin{split}
			I=\frac{e}{2h} \Gamma \tilde \Gamma \sum_{l=-\infty}^{\infty} \mathcal{L}_l \int &d\varepsilon \big[f_{L}^{e}(\varepsilon) - f_{R}^{e}(\varepsilon)\big]\big\{ (|\tilde{G}_{d11}^{r}(\varepsilon + l\omega_0)|^{2} + |\tilde{G}_{d12}^{r}(\varepsilon + l\omega_0)|^{2}) \big[f_{L}^{e}(\varepsilon + l\omega_0) + f_{R}^{e}(\varepsilon + l\omega_0)\big]\\
			&+ (|\tilde{G}_{d11}^{r}(\varepsilon - l\omega_0)|^{2} + |\tilde{G}_{d12}^{r}(\varepsilon - l\omega_0)|^{2}) \big[2 - f_{L}^{e}(\varepsilon - l\omega_0) - f_{R}^{e}(\varepsilon - l\omega_0)\big] \big\}.\\
		\end{split}
		\label{eq11}
	\end{equation}
\end{widetext}
The~dressed retarded Green's functions of the QD, $\tilde{G}_{d11}^r(\varepsilon)=\langle \langle d|d^{\dagger}\rangle \rangle_\varepsilon^r$ and $\tilde{G}_{d12}^r(\varepsilon)=\langle \langle d|d\rangle \rangle_\varepsilon^r$, in Eq.~\eqref{eq11} are calculated by employing the equation of motion technique~\cite{Mathe2020,Mathe2022}:
\begin{equation}
	\begin{aligned}
		&\tilde{G}_{d11}^{r}(\varepsilon)=\frac{\varepsilon + \tilde \varepsilon_d + i \tilde \Gamma - |\tilde\lambda|^2 K}{(\varepsilon - \tilde \varepsilon_d + i \tilde \Gamma)(\varepsilon + \tilde \varepsilon_d + i \tilde \Gamma) - 2(\varepsilon + i \tilde \Gamma)|\tilde\lambda|^2 K},\\
		&\tilde{G}_{d12}^{r}(\varepsilon)=\frac{-|\tilde\lambda|^2 K}{(\varepsilon - \tilde \varepsilon_d + i \tilde \Gamma)(\varepsilon + \tilde \varepsilon_d + i \tilde \Gamma) - 2(\varepsilon + i \tilde \Gamma)|\tilde\lambda|^2 K},
	\end{aligned}
	\label{eq12}
\end{equation}
with $K=\varepsilon/(\varepsilon^2 - \varepsilon_{M}^2)$. Note that if $\varepsilon_M = 0$, the~retarded Green's functions given by Eq.~\eqref{eq12} reduce to the results of~\cite{Wang2020}.
Note that the current can be simply determined at zero temperature by replacing the Fermi--Dirac function $f^{e}_\gamma(x)$ in Eq.~\eqref{eq11} with the Heaviside one $\theta(\mu_\gamma-x)$.
	
\section{Results and discussion}
\label{sec:III}
In the following, we discuss the transport properties of the QD--MBS system introduced above for a few experimentally relevant parameter~regimes.

As already mentioned in Section~\ref{sec:II}, the~system parameters must be smaller than the SC energy gap $\Delta$, which is on the order of $250 \,\mu eV$ in TSNW experiments~\cite{Mourik2012}. In~addition, our phonon-assisted transport calculations are performed in the limit where the QD--lead and QD--MBS couplings are weaker than the electron--phonon coupling strength (\mbox{$\Gamma,\,|\lambda|<\beta$})~\cite{Mathe2022}. In~our calculations, the~symmetrical QD--lead coupling $\Gamma$ is used as the energy unit. The~optical phonon energy $\omega_0$ and EPI strength are considered to be $\omega_0 = 5 \,\Gamma$ and $\beta = 2.5\, \Gamma$. For~more details regarding  the choice of parameters based on experimental measurements, see Ref.~\cite{Mathe2022} and references~therein.

We analyze the transport characteristics of our system at a finite temperature, thus serving as a relevant case for real systems. We also consider the $\beta$ and $T$ dependence of $\tilde \Gamma = \Gamma e^{-g(2N_{ph}+1)}$ and $|\tilde \lambda| = |\lambda|e^{-g(N_{ph}+1/2)}$.
Recall that the QD energy can be tuned by the gate voltage ($V_g$) applied to the gate electrode, i.e.,~$\varepsilon_d \propto V_g$.
In the following subsections, we show how the transport current is influenced by the system~parameters.\\

\begin{itemize}
	\item[\emph{(i)}]
	\textit{The effect of QD--MBS coupling $|\lambda|$ and bias voltage $V$ on current vs. gate voltage characteristics}
\end{itemize}

\begin{figure}[ht]
	\includegraphics[width =0.98\linewidth]{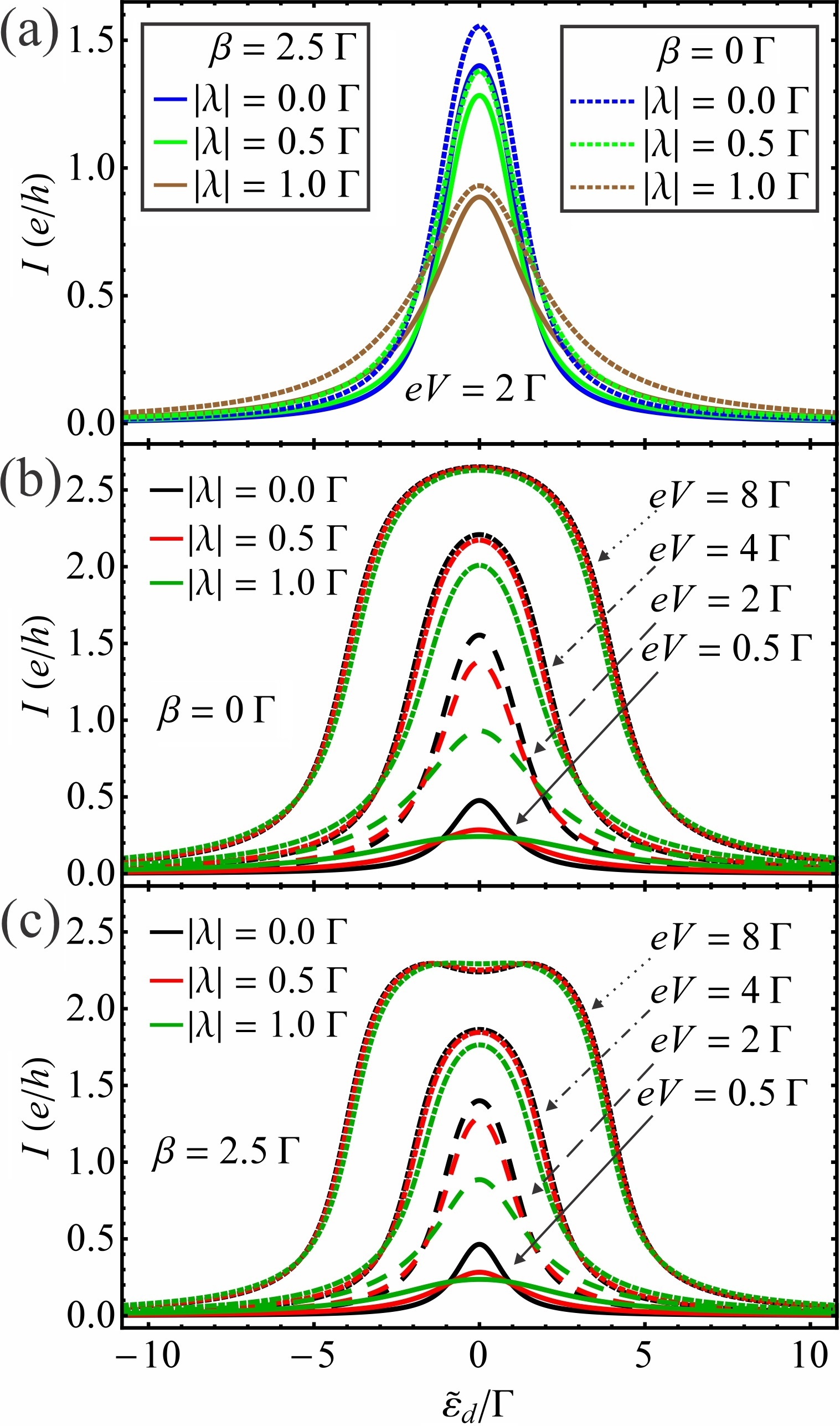}
	\centering
	\caption{(\textbf{a}) The current $I$ as a function of $\tilde\varepsilon_d$ for different values of the QD--MBS coupling $|\lambda|$ with unhybridized MBSs ($\varepsilon_M = 0\, \Gamma$). The~bias voltage is fixed as $eV = 2 \, \Gamma$, while the solid (dotted) lines correspond to the $\beta = 2.5 \, \Gamma$ ($\beta = 0 \, \Gamma$) case. The~current $I$ as a function of $\tilde\varepsilon_d$ for unhybridized MBSs ($\varepsilon_M = 0\, \Gamma$) at different values of bias voltage $eV$ and QD--MBS coupling $|\lambda|$ in the (\textbf{b}) absence and (\textbf{c}) presence of EPI with $\beta = 2.5\,\Gamma$. Here, the~solid, dashed, dot-dashed and dotted lines correspond to voltages $eV$ equal to $0.5\,\Gamma$, $2\,\Gamma$, $4\,\Gamma$ and $8\,\Gamma$, respectively. In~all cases, the temperature is fixed at $T = 0.1 \,\Gamma$.}
	\label{fig2}
\end{figure}

We first study the effect of QD--MBS coupling $|\lambda|$ and bias voltage $eV$ on the characteristics of current vs. $\tilde \varepsilon_d = \varepsilon_d - \beta^2/\omega_0$ for unhybridized Majoranas in the presence of EPI at a finite temperature $T=0.1\,\Gamma$. Figure~\ref{fig2}a shows the current $I$ as a function of $\tilde \varepsilon_d$ for different values of the dot--MBS coupling $|\lambda|$ when the system is biased as $eV = 2\, \Gamma$. The~calculations are made for unhybridized MBSs ($\varepsilon_M = 0$) in the absence and presence of EPI with a fixed electron--phonon coupling strength $\beta = 2.5\,\Gamma$.
We observe that in absence of EPI and MBSs, a~single Lorentzian resonant peak emerges at $\tilde \varepsilon_d = \varepsilon_d = 0$. The~amplitude of this peak is reduced when the dot couples to one of the ends of a TSNW (i.e., $|\lambda|\neq 0$) in the dot energy range $-(\Gamma + \frac{|eV|}{2}) \lesssim \varepsilon_d \lesssim (\Gamma + \frac{|eV|}{2})$. Beside the dot energy domain $|\varepsilon_d| \gtrsim (\Gamma + \frac{|eV|}{2})$, the~magnitude of the current $|I|$ increases slightly with $|\lambda|$ (see Figure~\ref{fig2}a, dotted lines). 
In the presence of EPI, the~current has a maximum at $\tilde \varepsilon_d =0$ and the absolute value of its amplitude is reduced compared to the $\beta = 0$ case. 
When the QD hybridizes with the MBS ($|\lambda| \neq 0$), the~spectrum of $I$ is changed (see Figure~\ref{fig2}a, solid lines). 
Similarly to the no EPI case, the~suppression of $|I|$ is realized when the renormalized dot energy $\tilde \varepsilon_d$ is situated within the energy region $-(\tilde \Gamma + \frac{|eV|}{2}) \lesssim \tilde \varepsilon_d \lesssim (\tilde \Gamma + \frac{|eV|}{2})$.
Otherwise, when $|\tilde\varepsilon_d| \gtrsim (\tilde\Gamma + \frac{|eV|}{2})$, a~slight increase in the current magnitude $|I|$ with $|\lambda|$ is observed.
This behavior of the current agrees qualitatively with the results of Ref.~\cite{Wang2021b} where the effect of EPI and temperature on $\Gamma$ and $|\lambda|$ is neglected by considering the couplings as constants.
The effect of the bias voltage $eV$ on the characteristics of $I-\tilde\varepsilon_d$ is shown without EPI in Figure~\ref{fig2}b and~with EPI of strength $\beta = 2.5\,\Gamma$ in Figure~\ref{fig2}c. 
In the absence of EPI and MBSs (Figure~\ref{fig2}b, black lines), 
the~magnitude of $|I|$ increases with the increase in bias $|eV|$, in~agreement with the literature~\cite{Sun2007}.
However, when the MBS is introduced in the system with $\beta = 0$ (see Figure~\ref{fig2}b, red and green lines), 
the~magnitude of the current is reduced with respect to the case of $|\lambda|= 0$ within the dot energy region $-(\Gamma + \frac{|eV|}{2}) \lesssim \varepsilon_d \lesssim (\Gamma + \frac{|eV|}{2})$, in~agreement with the result shown in Figure~\ref{fig2}a.
In the presence of EPI with MBSs (see Figure~\ref{fig2}c), 
the~current responds in the same way to the change in $|\lambda|$ for small values of the voltage $|eV|$ as in the $\beta = 0$ case under the mappings $\varepsilon_d \to \tilde \varepsilon_d$ and $\Gamma \to \tilde \Gamma$, respectively.
In the $\beta \neq 0$ case, the~effect of the hybridization of MBS with QD on the $I-\tilde \varepsilon_d$ curves alters depending on the bias voltage (discussed also in Figure~\ref{fig5}a below).
Consequently, the~change in current magnitude due to the QD--Majorana coupling $|\lambda|$ can be counteracted by tuning the bias~voltage.\\

\begin{figure}[ht]
	\includegraphics[width =0.98\linewidth]{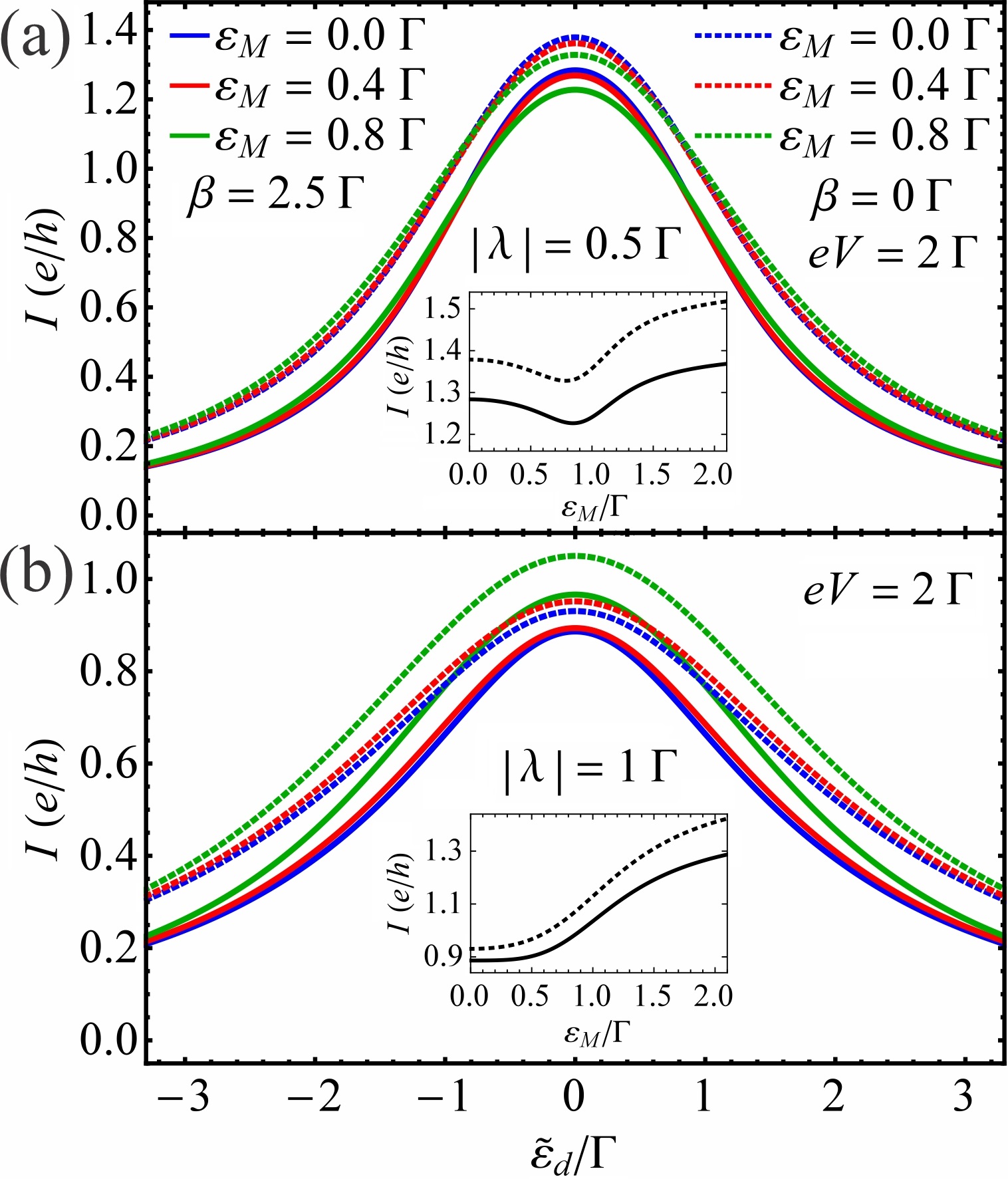}
	\centering
	\caption{The current $I$ as a function of $\tilde\varepsilon_d$ for different values of the overlap energy $\varepsilon_M$ with fixed electron--phonon coupling strength $\beta =2.5\,\Gamma$ at two values of the QD--MBS coupling $|\lambda|$: (\textbf{a})~$|\lambda| = 0.5\,\Gamma$ and (\textbf{b}) $|\lambda|=1\,\Gamma$. The~solid (dotted) lines correspond to the EPI presence (absence) case. Insets show current $I$ as a function of overlap energy $\varepsilon_M$ at $\tilde \varepsilon_d = 0\,\Gamma$ in the presence of EPI with $\beta =2.5\,\Gamma$ (solid line) and at $\varepsilon_d = 0$ in the absence of it (dotted line) for QD--MBS couplings: (\textbf{a}) $|\lambda|=0.5\,\Gamma$ and (\textbf{b}) $|\lambda|=1\,\Gamma$. The~temperature and bias voltage are $T=0.1\,\Gamma$ and $eV = 2\,\Gamma$, respectively.}
	\label{fig3}
\end{figure}

\begin{figure*}[ht]
	\includegraphics[width =0.85\linewidth]{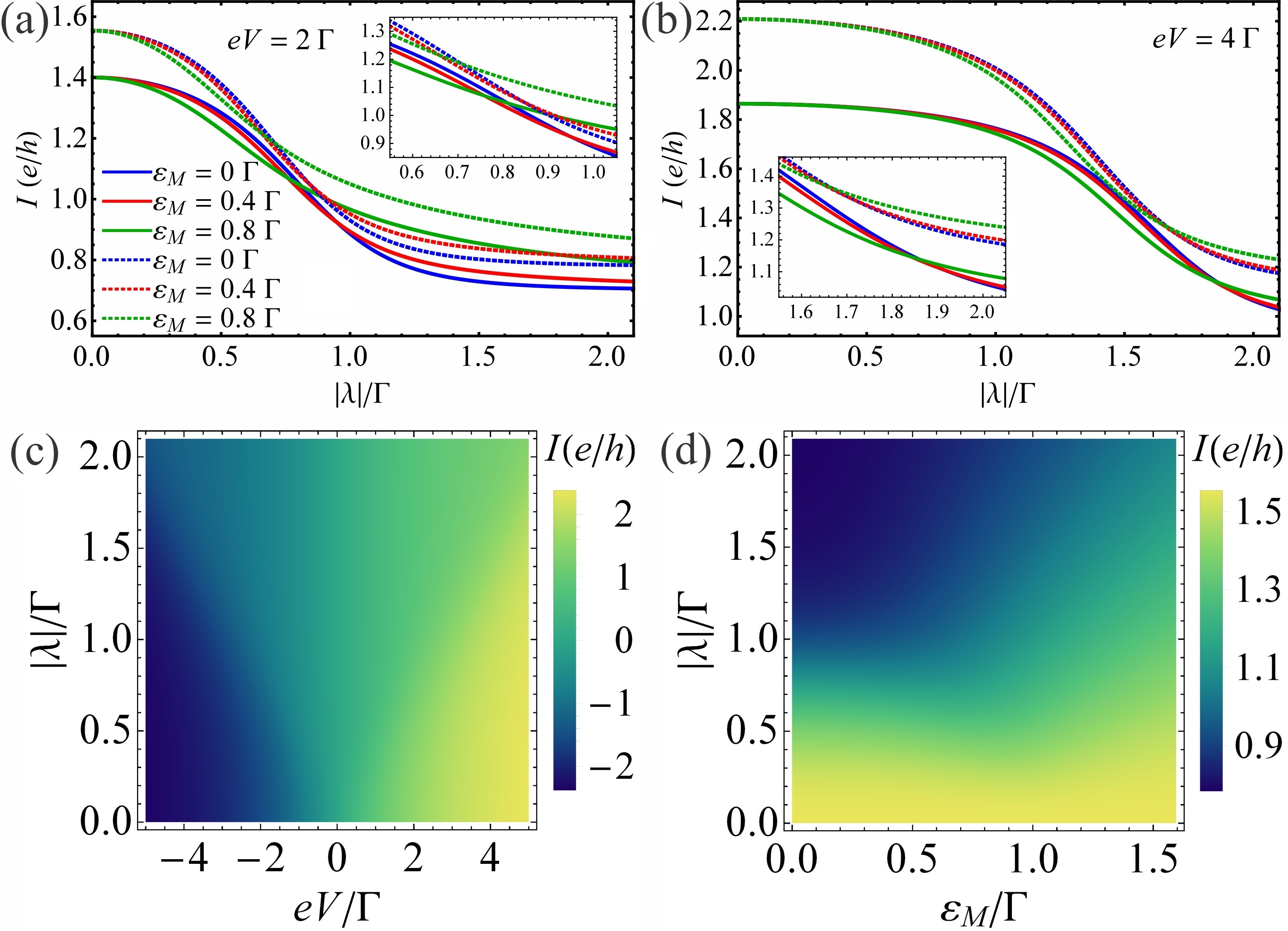}
	\centering
	\caption{The current $I$ as a function of QD--MBS coupling $|\lambda|$ at different values of the overlap energy $\varepsilon_M$ for bias voltage (\textbf{a}) $eV =2\,\Gamma$ and (\textbf{b}) $eV = 4\,\Gamma$. The~solid (dotted) lines correspond to the $\beta = 2.5\,\Gamma$ at $\tilde \varepsilon_d =0\,\Gamma$ ($\beta = 0\,\Gamma$ at $\varepsilon_d = 0 \,\Gamma$) case. The~insets in each panel zoom in on current.	
	(\textbf{c}) The current $I$ as a function of bias voltage $eV$ and QD--MBS coupling $|\lambda|$ for unhybridized MBSs ($\varepsilon_M = 0\,\Gamma$) in the absence of EPI. (\textbf{d}) The current $I$ as a function of overlap energy $\varepsilon_M$ and QD--MBS coupling $|\lambda|$ at bias voltage $eV = 2\,\Gamma$ in the absence of EPI. The~QD energy is $\varepsilon_d = 0\,\Gamma$ in (\textbf{c},\textbf{d}). The~considered temperature is $T = 0.1\,\Gamma$.}
	\label{fig4}
\end{figure*}

\begin{itemize}
	\item[\emph{(ii)}]
	\textit{The effect of MBS overlap energy $\varepsilon_M$ on current vs. gate voltage characteristics}
\end{itemize}

Next, we investigate the effect of the overlap energy $\varepsilon_M$ on the $I-\tilde\varepsilon_d$ characteristics. Figure~\ref{fig3} shows the results for the current $I$ as a function of QD energy $\varepsilon_d$ in the absence of EPI and as a function of $\tilde \varepsilon_d$ in the presence of EPI with coupling $\beta = 2.5 \,\Gamma$ at different values of the overlap energy $\varepsilon_M$ for two QD--MBS coupling $|\lambda|$ values. The~temperature and bias voltage are fixed at $T = 0.1\,\Gamma$ and $eV = 2\,\Gamma$, respectively. We observe that the current magnitude around  $\tilde \varepsilon_d = \varepsilon_d = 0$ without EPI and  $\tilde\varepsilon_d = 0$ with EPI reduces with the increase of overlap energy $\varepsilon_M$ when the QD weakly couples to the MBS (see Figure~\ref{fig3}a). 
In the weak $|\lambda|$ case with a given $\varepsilon_M \neq 0$, we also see that by moving away from $\varepsilon_d =0$ for $\beta = 0$ (or $\tilde\varepsilon_d =0$ for $\beta \neq 0$) to larger $|\varepsilon_d|$ (or $|\tilde\varepsilon_d|$) values, there is a critical value above which the current changes from a reduction to an enhancement relative to the $\varepsilon_M=0$ curve.
Therefore, we observe that a further increase in $\varepsilon_M$ leads to the current amplitude at $\tilde \varepsilon_d = 0$ with EPI (or at $\varepsilon_d = 0$ without EPI) beginning to increase (see the inset in Figure~\ref{fig3}a). 
At stronger $|\lambda|$ (see Figure~\ref{fig3}b with its inset), the~current reacts differently to the change in $\varepsilon_M$ relative to the $|\lambda| = 0.5\,\Gamma$ case.
Namely, near~$\varepsilon_d = 0$ for $\beta = 0$ or $\tilde \varepsilon_d = 0$ for $\beta \neq 0$, the~current amplitude increases with $\varepsilon_M$.
When the QD level passes a critical energy value, the~current magnitude at a given $\varepsilon_M$ will be reduced relative to the $\varepsilon_M = 0$ case.
This critical dot energy value moves to smaller $|\varepsilon_d|$ with the increase in $\varepsilon_M$.
Note that the current curve peak in the presence of EPI (Figure~\ref{fig3}, solid lines) is narrower than the one without EPI (Figure~\ref{fig3}, dotted lines) because of the renormalized QD--lead coupling $\tilde \Gamma$.

To further understand the regime $eV \neq 0$ and $|\lambda|\neq 0$ with $\varepsilon_M\neq 0$, we plot in Figure~\ref{fig4}a,b the current $I$ for different values of the overlap energy $\varepsilon_M$ and bias voltage $eV$ in the absence of EPI at $\varepsilon_d = 0$ and in the presence of EPI at $\tilde \varepsilon_d = 0$ with electron--phonon coupling strength $\beta = 2.5\,\Gamma$, respectively. These results are represented as a function of QD--MBS coupling $|\lambda|$ at a finite temperature $T =0.1\,\Gamma$.
We can see that in the absence of EPI with unhybridized Majoranas (see Figure~\ref{fig4}a,b, blue dotted lines), the~maximum of the current that emerges at $\varepsilon_d = 0$ significantly reduces when the dot hybridizes with the MBS, in~agreement with the result from Figure~\ref{fig2}a. We can observe that $|I|$ increases with $|eV|$ at fixed values of $|\lambda|$, in~agreement with the results from Figure~\ref{fig2}b.
Therefore, the~value of QD--MBS coupling $|\lambda|$ where $|I|$ presents a significant reduction, shifts to higher values of $|\lambda|$ with the increase in bias $|eV|$. 
In the case of hybridized MBSs, at~a given $\varepsilon_M$, there is a critical value for QD--MBS coupling, where the effect of $\varepsilon_M$ on the current amplitude changes from a reduction to an enhancement relative to the $\varepsilon_M=0$ case.
This critical value of $|\lambda|$ depends on the bias voltage $eV$.
The increasing bias $|eV|$ leads to negligible influence of the finite overlap energies considered here on the current curves.
Note here that a relation for the current $I(eV) = - I(-eV)$ can be established, as~seen also in Figure~\ref{fig5} below. 
To better understand the current-bias voltage dependence when the dot--Majorana coupling changes, we plot in Figure~\ref{fig4}c the current $I$ as a function of $eV$ and $|\lambda|$ in the absence of EPI for unhybridized MBSs at $\varepsilon_d = 0$ and a finite temperature $T = 0.1\,\Gamma$. 
In Figure~\ref{fig4}c, we notice that the line $|\lambda|\propto eV$ represents an inflection point which corresponds to a peak in the differential conductance.
Figure~\ref{fig4}d shows the results for the current $I$ as a function of overlap energy $\varepsilon_M$ and QD--MBS coupling strength $|\lambda|$ in the absence of EPI at a fixed voltage $eV = 2\,\Gamma$, dot energy $\varepsilon_d = 0$ and temperature $T = 0.1\,\Gamma$.
The current map details further changes with the enhancement of the MBS--MBS coupling strength $\varepsilon_M$ predominantly when the QD strongly couples to the MBS (see also Figure~\ref{fig4}a).
In the presence of EPI (see Figure~\ref{fig4}a,b, solid lines), the~amplitude of current $|I|$ is reduced relative to the $\beta = 0$ case which is more visible at voltages near $eV \approx 4\,\Gamma$ (see also Figure~\ref{fig5} below for a larger $eV$ domain).
The $I-|\lambda|$ curves for $\beta \neq 0$ show the same behavior as those for $\beta = 0$ at low bias~voltages.\\

\begin{itemize}
	\item[\emph{(iii)}]
	\textit{The effect of QD--MBS coupling $|\lambda|$ and MBS overlap energy $\varepsilon_M$ on current vs. bias voltage characteristics}
\end{itemize}

In the following, we investigate the current-bias voltage characteristics of the proposed system at different values of the QD--MBS coupling $|\lambda|$ in the absence and presence of EPI for both unhybridized and hybridized MBSs. The~results are shown in Figure~\ref{fig5} at $\tilde \varepsilon_d = 0$ with EPI of strength $\beta = 2.5 \,\Gamma$ and at $\varepsilon_d=0$ without EPI and~at a finite temperature $T=0.1\,\Gamma$. The~current $I$ shows a step-like structure as a function of bias voltage $eV$. This structure is explained below. 
In the absence of EPI and MBSs (see Figure~\ref{fig5}a, red dotted line), when the system is positively biased ($eV>0$) and $\mu_R < \varepsilon_d < \mu_L$ (with $\mu_L = - \mu_R = eV/2$), the~dot is able to receive an electron from lead $L$ and transfer it to lead $R$, which results in a current passing through the dot. Such a system (without MBS or EPI) has already been detailed in Ref.~\cite{Zimbovskaya2008}.
When the dot couples to the MBS, the~magnitude of the current $|I|$ reduces in the vicinity of zero-bias voltage with the increase in QD--MBS coupling $|\lambda|$ for unhybridized MBSs (see Figure~\ref{fig5}a with its inset, dotted lines). The~width of this voltage window, where the current is affected by the QD--MBS coupling, becomes larger on enhancing $|\lambda|$, in~agreement with the findings from Figure~\ref{fig4}c.
In the presence of EPI and absence of MBSs (see Figure~\ref{fig5}a, black solid line), the~$I-V$ curve is visibly different and new steps show up in the spectrum which correspond to the opening of phonon-assisted tunneling channels~\cite{Zhou2019}.
When the dot hybridizes with the MBS in the $\beta \neq 0$ case for $\varepsilon_M = 0$ (see Figure~\ref{fig5}a, blue and green solid lines), further changes in the $I-V$ characteristics can be observed.
For instance, in~the positive bias domain, i.e.,~$eV \gtrsim 0$, we see that the amplitude of $I$ decreases with the enhancement of QD--MBS coupling $|\lambda|$ when the bias voltage is approximately within $2l\omega_0 \lesssim eV \lesssim (2l+1)\omega_0$. Instead, this amplitude increases with $|\lambda|$ in the bias regimes $(2l+1)\omega_0 \lesssim eV \lesssim 2(l+1)\omega_0$ with $l = 0,1,2, \dots$. These findings are consistent with the results reported in Ref.~\cite{Wang2021b}. 
For negative voltages ($eV \lesssim 0$), the~changes in current are similar, taking into account the antisymmetric nature of the $I-V$ curves. 
Thus, the~current $|I|$ decreases with the increase in $|\lambda|$ when $eV$ is within \linebreak \mbox{$(2l+1)\omega_0 \lesssim eV \lesssim 2(l+1)\omega_0$} and it increases with $|\lambda|$ when $2l\omega_0 \lesssim eV \lesssim (2l+1)\omega_0$ with $l = -1, -2,\dots $, respectively.

\begin{figure}[ht]
	\includegraphics[width =0.98\linewidth]{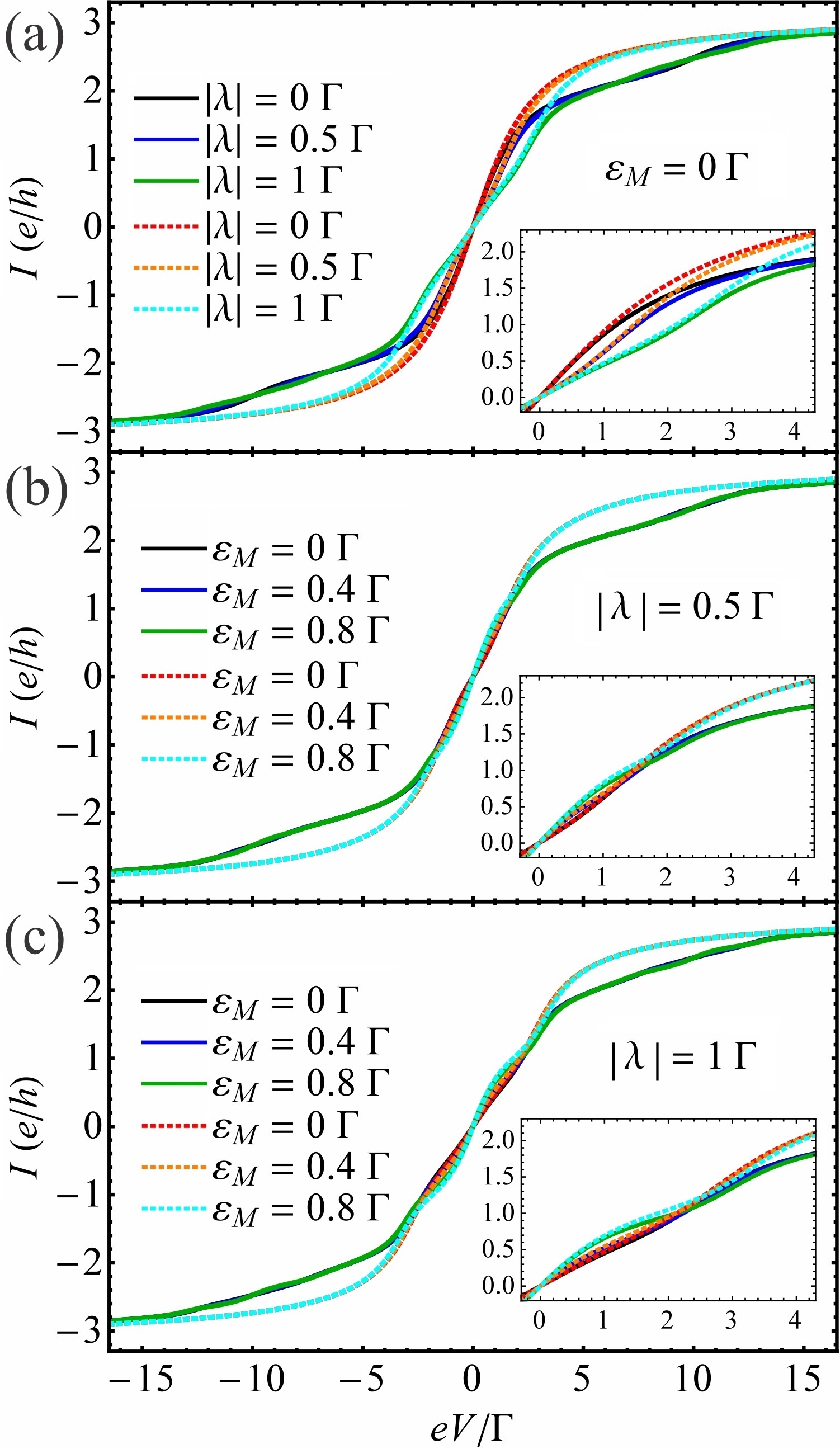}
	\centering
	\caption{(\textbf{a}) The current $I$ as a function of bias voltage $eV$ at different values of QD--MBS coupling $|\lambda|$ for unhybridized MBSs ($\varepsilon_M = 0\, \Gamma$). The~current $I$ as a function of bias voltage $eV$ at different values of the overlap energy $\varepsilon_M$ for QD--MBS couplings: (\textbf{b}) $|\lambda| = 0.5 \,\Gamma$ and (\textbf{c}) $|\lambda| = 1 \,\Gamma$, respectively. The~solid and dotted lines correspond to the presence and absence of EPI. The~electron--phonon coupling strength and renormalized dot energy are $\beta = 2.5 \, \Gamma$ and $\tilde \varepsilon_d = 0 \, \Gamma$, respectively. In~the absence of EPI the QD energy is $\varepsilon_d=0\,\Gamma$. The~temperature is $T = 0.1\, \Gamma$. The~insets in each panel zoom in on current near the zero-bias at positive~voltage.}
	\label{fig5}
\end{figure}

The effect of overlap energy $\varepsilon_M$ on current-bias voltage characteristics is illustrated in Figure~\ref{fig5}b,c for two values of the dot--MBS coupling $|\lambda|$, both in the absence at $\varepsilon_d = 0$ and presence of EPI with strength $\beta = 2.5 \, \Gamma$ at $\tilde \varepsilon_d = 0$.
In the absence of EPI (see Figure~\ref{fig5}b,c, dotted lines), we observe that the current magnitude $|I|$ increases with the overlap energy $\varepsilon_M$ near the zero-bias regime, when the voltage is constrained within $|eV| \lesssim \mathcal{V}$. Here, $\mathcal{V} \propto (2|\lambda| + \varepsilon_M)$ for a given $|\lambda| \neq 0$.
Otherwise ($|eV|\gtrsim \mathcal{V}$), the~current $|I|$ reduces slightly on increasing $\varepsilon_M$.
In the presence of EPI for hybridized Majoranas (see Figure~\ref{fig5}b,c, solid lines), the~finite $\varepsilon_M$ significantly influences the current-bias voltage characteristics. Namely, for~a positively biased QD system ($eV \gtrsim 0$), the~magnitude of the current $|I|$ increases with $\varepsilon_M$ when the bias voltage is located within $2l\omega_0 \lesssim eV \lesssim 2l\omega_0 +\mathcal{\tilde V}$ and $(2l+1)\omega_0 \lesssim eV \lesssim 2(l+1)\omega_0 - \mathcal{\tilde V}$ and~decreases with the increase in $\varepsilon_M$ when $eV$ is within $2l\omega_0 +\mathcal{\tilde V}\lesssim eV \lesssim (2l+1)\omega_0$ and $2(l+1)\omega_0 -\mathcal{\tilde V} \lesssim eV \lesssim 2(l+1)\omega_0$, with~$l=0,1,2,\dots $.
Here, we introduced the notation $\mathcal{\tilde V} \propto (2|\tilde\lambda| + \varepsilon_M)$ for a given $|\tilde \lambda| \neq 0$ with the restriction $0<\mathcal{\tilde V}< \omega_0$.
For a negatively biased QD ($eV \lesssim 0$), the~alteration of the current behavior as a response to the change in $\varepsilon_M$ is expressed similarly to the case of $eV \gtrsim 0$ by taking into account the asymmetric property of $I-V$ characteristics.
Consequently, when the dot couples to the MBS, the~current can be amplified or reduced by changing the value of Majorana overlap energy at a fixed bias~voltage.\\

\begin{figure}[ht]
	\includegraphics[width =0.98\linewidth]{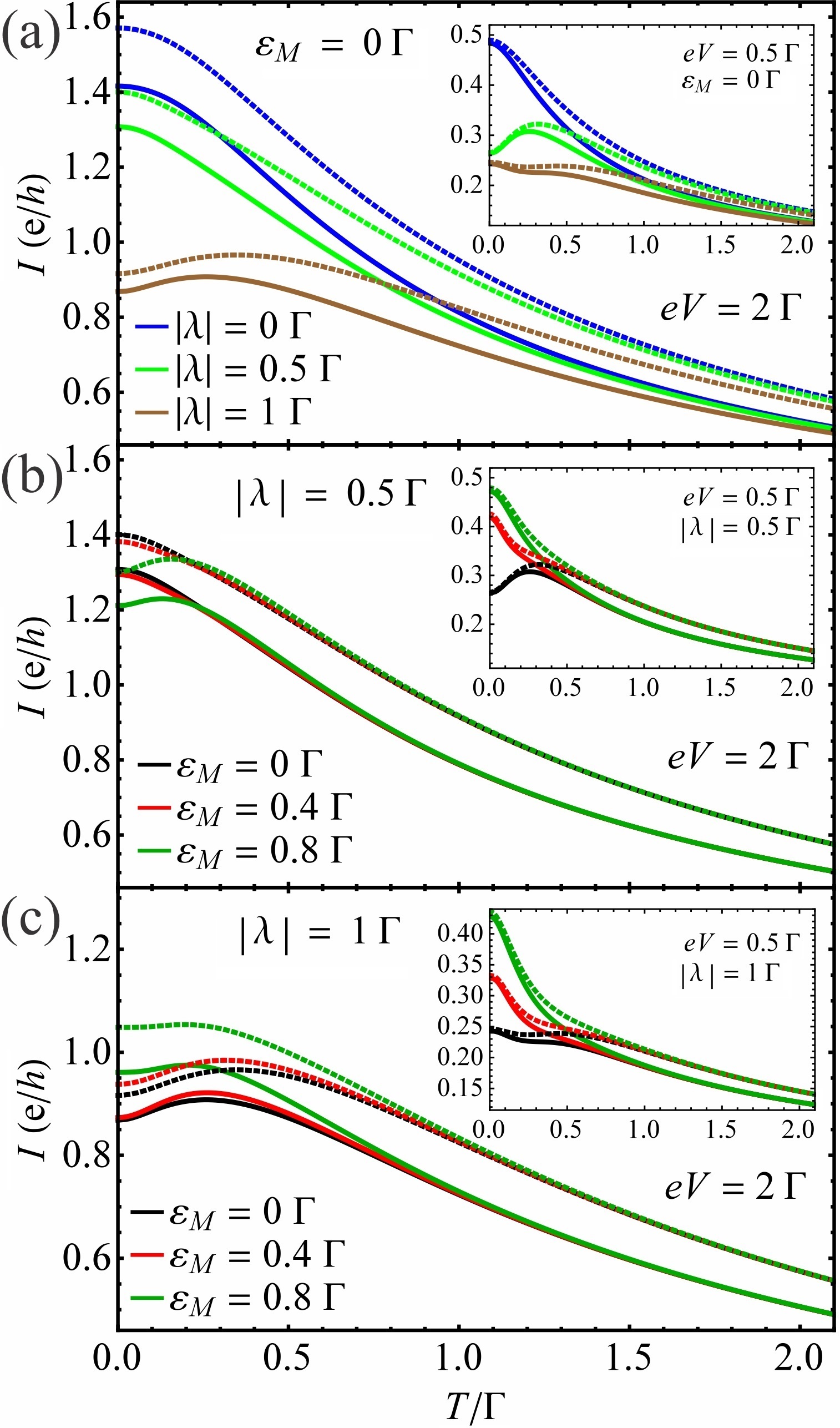}
	\centering
	\caption{(\textbf{a}) The current $I$ as a function of temperature $T$ at different values of the QD--MBS coupling $|\lambda|$ for unhybridized MBSs ($\varepsilon_M = 0\,\Gamma$) at bias voltage $eV = 2\,\Gamma$. The~inset in (\textbf{a}) shows the results at $eV = 0.5\,\Gamma$. The~current $I$ as a function of temperature $T$ at different values of the overlap energy $\varepsilon_M$ and at bias voltage $eV = 2\,\Gamma$ for QD--MBS couplings: (\textbf{b}) $|\lambda| = 0.5\,\Gamma$ and (\textbf{c}) $|\lambda| = 1\,\Gamma$. The~insets in (\textbf{b},\textbf{c})  show the results at bias $eV = 0.5 \,\Gamma$. The~solid (dotted) lines represent the results with (without) EPI. The~renormalized QD energy is $\tilde \varepsilon_d =0 \,\Gamma$ in the presence of EPI while the dot energy is $\varepsilon_d = 0\,\Gamma$ in its absence. The~electron--phonon coupling strength is $\beta= 2.5\,\Gamma$.}
	\label{fig6}
\end{figure}

\begin{itemize}
	\item[\emph{(iv)}]
	\textit{The effect of dot--MBS coupling $|\lambda|$ and Majorana overlap energy $\varepsilon_M$ on current vs. temperature characteristics}
\end{itemize}

We plot in Figure~\ref{fig6} the current $I$ as a function of temperature $T$ for different values of the bias $eV$, QD--MBS coupling $|\lambda|$ and overlap energy $\varepsilon_M$ in the absence at $\varepsilon_d = 0$ and in the presence of EPI with $\beta=2.5\,\Gamma$ at $\tilde \varepsilon_d=0$. We first observe that the current is suppressed for $\beta \neq 0$ relative to its value at $\beta = 0$ (see Figure~\ref{fig6}a--c and insets as well as solid and dotted lines). 
We find that the current shows a nonmonotonic behavior with temperature $T$ for unhybridized Majoranas when the QD weakly couples to the MBS at low bias voltage in both the $\beta \neq 0$ and $\beta = 0$ cases (inset in Figure~\ref{fig6}a, green solid and dotted lines). Namely, $|I|$ increases with $T$ up to a specific value and~begins decreasing above that value.
With the increase in $|\lambda|$ this tendency of the $I-T$ characteristics starts to vanish (inset in Figure~\ref{fig6}a, brown solid and dotted lines).
As the bias increases, the~nonmonotonic behavior of the current with $T$ emerges at stronger coupling $|\lambda|$ (see Figure~\ref{fig6}a, brown solid and dotted lines).
Note here that this tendency of the $I-T$ characteristics disappears at higher voltages for values of $|\lambda|$ which satisfy the approximation applied here.
In the case of hybridized MBSs with relatively strong overlap energy, the~nonmonotonic behavior of the current with temperature disappears at small biases $eV$ (see the inset in Figure~\ref{fig6}b, red and green lines).
In addition, $|I|$ varies nonmonotonically with temperature at weaker $|\lambda|$ with stronger $\varepsilon_M$ at bias $eV = 2\,\Gamma$ (Figure~\ref{fig6}b, green solid and dotted lines).
This behavior of the $I-T$ curves starts to vanish with further increase in $\varepsilon_M$.
When the QD couples strongly to the MBS, the~current decreases with temperature at low bias voltages ($eV = 0.5\,\Gamma$, see the inset in Figure~\ref{fig6}c). 
The nonmonotonic behavior of the current--temperature curves holds for values of $\varepsilon_M$ considered here at $eV = 2\,\Gamma$ and $|\lambda|=1\,\Gamma$ (see Figure~\ref{fig6}c). 
Similarly to the $|\lambda| = 0.5\,\Gamma$ case, the~further increase in $\varepsilon_M$ smears the nonmonotonic behavior of the current as a function of temperature at $eV = 2\,\Gamma$.
Note here that the nonmonotonic behavior of current with temperature vanishes at higher bias voltages regardless of the $|\lambda|$ coupling values within the limits set by the approximations used here. In~this case, the~overlap energy also shows less influence on the $I-T$ characteristics.
Consequently, the~response of the current to the changes in temperature is altered nontrivially depending on bias voltage, QD--MBS coupling and Majorana overlap~energy.

\section{Conclusions}
\label{sec:IV}
\setcounter{equation}{0}

In the present work, we have investigated the complex physics of a QD coupled to a MBS located at one of the edges of a TSNW. In~addition, the~phonon-assisted transport properties of the considered setup were explored in the subgap regime when the localized electrons in the QD interact with a single long-wave optical phonon mode.	
When determining the current, the~EPI was treated by employing a canonical transformation within the nonequilibrium Green's function technique.
We discussed in detail the effect of EPI on the current vs. gate voltage, current vs. bias voltage and current vs. dot--Majorana coupling characteristics for unhybridized and hybridized MBSs at finite temperature.
We established that in the absence of EPI the dot--Majorana coupling strength suppresses the current when the dot energy is located near the Fermi level, especially at low bias voltages; therefore, the Majorana overlap energy and dot--Majorana coupling have a more significant impact on the transport under this low bias regime.
The effect can be counteracted by increasing the bias voltage.
In the presence of EPI, the~effect of dot--MBS coupling on the current-gate voltage characteristics can be regulated by changing the bias voltage.
The current-bias voltage curves present a step-like structure in the presence of electron--phonon coupling due to the phonon-assisted tunneling through the dot.
The effect of Majorana overlap energy on current vs. bias voltage characteristics alters depending on the bias voltage value.
Note that the current is insensitive to charge fluctuations at high voltages.
We found that the current shows a nonmonotonic behavior with temperature depending on the values of QD--MBS coupling, overlap energy, gate and bias voltages.
The current shows sizable changes at low temperatures if the bias voltage is low. Note that the current--temperature dependence is strongly affected even by lower gate voltage variations.
In the future, we plan to extend this investigation to determine if such systems possess parameter regimes for which it is easy to establish the presence or the absence of MBSs in the system via transport measurements.
Finally, the~device geometry investigated in this work should be experimentally realizable by taking into consideration the recent advancements in the field~\cite{Deng2016,Deng2018,Razmadze2020}. Our investigation can serve as a guide for experiments probing MBSs with QDs, helping to enlarge the understanding of topological quantum~computation.\\

\section*{Acknowledgments} 
The authors would like to thank  Doru Sticle\c{t},  Luiza Buimaga-Iarinca, Larisa-Milena Pioraș-Țimbolmaș and Pál-Attila Máthé for valuable~discussions. L.M. and L.P.Z. acknowledge financial support from the MCID through the ``Nucleu'' Program within the National Plan for Research, Development and Innovation 2022--2027, project PN~23~24~01~04 and through PNCDI III---Program 1---Development of the National Research and Development System, Subprogram 1.2---Institutional Performance---Funding Projects for Excellence in RDI, Contract No.~37PFE/30.12.2021.

\bibliographystyle{apsrev4-2}
\bibliography{ReferencesPhonon}
\end{document}